\documentclass[prb,twocolumn,showpacs,amsmath,amssymb]{revtex4}
\usepackage{bm}
\usepackage{graphics}
\begin{document}

\title{Nonergodic thermodynamics of disordered ferromagnets and ferroelectrics}
\author{P. N. Timonin}
\email{timonin@aaanet.ru}

\affiliation{Physics Research Institute at Southern Federal University,
344090, Rostov - on - Don, Russia}
 
\date{\today}

\begin{abstract}
Phenomenological thermodynamic theory describing the properties of metastable states in disordered ferromagnets and ferroelectrics with frustrative random interactions is developed and its ability to describe various nonergodic phenomena in real crystals is demonstrated. 
\end{abstract}


\maketitle
\section{Introduction}
\label{intro}
Many types of disorder in ferromagnets and ferroelectrics can induce spin- or dipole-glass states if emergent random interactions cause frustration and make the ground state of a crystal highly degenerate \cite{1}. Besides the classic example of the solid solution (alloy) of ferromagnet with antiferromagnet \cite{2} it can be the random anisotropy \cite{3}, simple dilution in the case of competing nearest and next-nearest interactions \cite{4} or in the presence of strong dipole-dipole ones \cite{5} as well as structural defects such as dislocations \cite{6}. Now it is widely recognized that the unusual nonergodic properties of disordered ferromagnets and ferroelectrics in their glassy phases result from the appearance of a number of metastable states each having more or less random directions of local spontaneous electric or magnetic dipole moments \cite{1}. Using the different protocols of changing temperature and external field (e. g. field-cooled, FC, zero-field cooled, ZFC, etc.) one may arrive at the metastable states with different net magnetization (polarization) and differences in other thermodynamic functions.

First adequate theoretical description of such history - dependence of system properties in glassy phases was achieved in Refs. [\onlinecite{7}, \onlinecite{8}] using numerical simulations in the local mean-field theory of random-bond magnets. It was explicitly shown that various regimes of temperature and field variations do bring system to different minima of thermodynamic potential thus providing the history - dependence of thermodynamic functions. Under the assumption of the possibility of sufficiently slow (quasi-static) variations of external parameters making the predictions of static thermodynamic theory valid the number of nonergodic features of glassy phases were described in Refs. [\onlinecite{7}, \onlinecite{8}]: the temperature dependencies of FC and ZFC functions, thermal and isothermal remanent magnetizations, the form of hysteresis loops, FC heat capacity in different fields. These results are in reasonable agreement with the experimental data for a series of random magnets. Later the method of the local mean-field simulations were used to reveal the presence of nonergodic reentrant spin-glass phase in 2d XY random-bond model \cite{9}.

Subsequent studies of nonergodic phenomena in spin systems with random interactions were mainly focused on numerical simulations of hysteresis loops. The smaller loops inside the main one and evolution via macroscopic spin avalanches on the main loop are shown to exist in Edwards-Anderson \cite{10} and Sherrington- Kirkpatrick \cite{11,12} spin-glass models. The existence of hysteresis loops is established in random - anisotropy model \cite{13}. The new hard-spin mean-field method is developed for frustrated random - bond systems \cite{14,15,16} which provides the evidences for the existence of multiple metastable states and can describe a number of nonergodic phenomena with less simulation efforts.

In spite of definite successes in numerical simulations of nonergodic effects in glassy phases these methods are not destined to provide some general picture for systematization and qualitative explanation of these effects. It is rather difficult to find from the existing numerical results the possible interrelations between a variety of history - dependent phenomena observed in different experimental regimes, see Refs. [\onlinecite{17}-\onlinecite{25}]. Yet it seems that they should be necessary present as, e. g., the hysteresis loops and ZFC - FC magnetization differences have the common origin. Then one may hope to obtain the unified picture of such phenomena using analytical calculations in some simplified models. But nowadays there are no simple enough microscopic models in which nonergodic effects in finite field could be described by the analytical methods.
 
Thus in present situation some phenomenological approach may help to achieve general understanding of the quasi-static large field response in glassy phases and the role played in it by metastable states. Apparently, it should be based on some realistic mechanism underlying the appearance of such states. Such mechanism for the random-interaction systems has been described in Ref. [\onlinecite{26}]. It consists in the subsequent condensations of numerous sparse fractal modes defined by the eigenvectors of the random matrix of pair-wise interactions (exchange matrix in magnets or the matrix of harmonic interaction of polar atomic displacements in ferroelectrics) at their localization threshold \cite{27}. In random ferroelectrics (relaxors) these modes play the role of the notorious "soft mode" as they are the delocalized eigenmodes of polar atomic oscillation with the lowest frequencies. Their amplitudes throughout the disordered crystal can be obtained by the numerical methods of Refs. [\onlinecite{27}]. The condensation (freezing) of these modes in relaxors results in appearance of stable polar atomic displacements same as in the case of ordinary soft mode albeit these displacements are highly inhomogeneous and sparse. Their analogs in random Ising magnets can be related to the delocalized relaxational magnetic modes which freeze first at a macroscopic spin glass transition. 

It is important to realize that only the freezing of these macroscopic modes can result in emergence of stable spontaneous local moments at a glass transition. Such objects as nano-domains will never have spontaneous moments stable for a macroscopic time so this notion is useless for the description of the true thermodynamic transition into the glass phase. Thus Burns temperature in relaxors at which experiments reveal the appearance of (fluctuating) polar nano-domains marks the onset of the paraelectric Griffiths' phase in which only non-analyticity in the field dependence of polarization \cite{28} and slow non-exponential relaxation \cite{29} appear.

The suggested mechanism of a macroscopic spin glass transition is in sharp contrast with that of phase transition in homogeneous crystals. The last takes place due to the condensation of just one periodic or uniform in space eigenmode which corresponds to the largest eigenvalue of exchange matrix (in magnets) or the lowest eigenvalue of the matrix of effective harmonic interaction (soft mode in ferroelectrics). In ideal crystal it is sufficient to stabilize the other modes with eigenvalues close to the critical one throughout some low-symmetry phase regions. This stabilization of the rest of near-critical modes results from nonlinear mode couplings which are always present if there is some finite spatial overlap between the critical mode and near-critical ones. As all eigenmodes of an ideal crystal is just the plane waves strongly overlapping throughout the whole sample, the condensation of just one critical mode makes the system stable.
 
In a strongly disordered crystal with random interactions the situation changes principally. The randomness of pair-wise interactions makes the eigenmodes near the boundary of the spectrum localized \cite{27} so the transition may start only with the condensation of the first delocalized eigenmode at the localization threshold somewhat far from the lower boundary of the spectrum. Owing to the sparse fractal structure of the modes near this threshold \cite{27}, the condensation of the first delocalized mode can not stabilize the other sparse near-critical modes which do not overlap essentially with the first one. So they proceed to condense at slightly lower temperatures until almost all crystal sites acquire spontaneous moments. If the average fractal dimension of these condensing modes is $d_f < d$ , the average number of sites participating in each mode is of the order  $N_1  = N^{d_f /d}$. So the number of them needed to cover the $N$-site crystal is $N_0  = N/N_1  = N^{1 - \left( {d_f /d} \right)} $. 

Thus strong disorder can transform the ordinary phase transition into the macroscopic number of them spreading in some temperature interval. As each mode after condensation can have (at least) two stable states with the reverse directions of local moments, we may end up with up to $2^{N_0 }$ metastable states. Note that each of these numerous subsequent transitions is still a macroscopic one as the macroscopic number of local spontaneous moments of order $N_1$ appears at it. Apparently, the mean-field estimate of the potential barrier between metastable states would also give the macroscopic value proportional to $N_1$.
 
Here we may note that above picture describes only the possibility of emergence of numerous metastable states in crystal with random interactions. Do it actually realizes depends on the details of pair-wise interaction between the sparse fractal modes. If this interaction tends to orient the net moments of the modes in parallel, we may have ordinary ferromagnetic (ferroelectric) phase with or without some metastable states. Principally, sketched above mechanism allows for phenomenological description of the thermodynamics of metastable states in the spirit of Landau's theory of phase transitions. The case with purely glass transition was considered in Ref. [\onlinecite{26}]. Here we present the simplified and more general derivation of the phenomenological Landau's potential for the generic case of competing ferromagnetic (ferroelectric) and glassy interactions in disordered crystals. For a large class of such random systems it is possible to obtain semi-analytical description of the properties of the emergent metastable states thus providing the comprehensive account of possible nonergodic effects in random ferromagnets and ferroelectric relaxors.

\section{Landau's potential for random ferromagnets and ferroelectrics}
\label{sec:1}
Further we mainly resort to the magnetic terminology and designations just to be specific. Let us consider the ferromagnet with the second-order transition and one-component order parameter where some disorder of random-interaction type presents tending to destroy the ferromagnetic order. According to above considerations we can introduce in this case the Landau potential for some specific realization of disorder depending on the net magnetizations of sparse fractal modes, $m_i, i = 1,...,N_0$. If the temperature interval ($T_g $ , $T_g  - \Delta T$) at which the condensation of these modes takes place is narrow,  $ \Delta T  \ll T_g $ then we can consider the small $m_i$ near $T_g$ and expand the potential $F\left( {\bf m} \right)$ in powers of $m_i$. It would contain only even powers of these order parameters due to the global inversion symmetry. Also we can retain in it the terms with no more than two different modes as one can suggest that their sparse spatial structure makes the simultaneous interactions of three or more modes negligibly small. 

Further we may note that $\Delta T  \ll T_g $  is actually the necessary condition when one consider the frustrating disorder causing the (near) degeneracy of condensing types of magnetic (dipole) arrangements. This means that in such case the condensing modes have nearly the same thermodynamic potentials, close transition temperatures and thermodynamic parameters. The simplest way to imitate this near degeneracy is to suppose that all thermodynamic properties of all modes are identical. Then $ F\left( {\bf m} \right)$  should be symmetric under all permutations of $m_i$. The immediate effect of this assumption will be the merging of actually subsequent transitions to just one point, $T_g$. In this way we avoid the consideration of narrow temperature interval ($T_g $ , $T_g  - \Delta T$) and gain the significant simplification of the model which still preserve the essential features of nonergodic transitions caused by frustrating disorder. We may say that model with permutation symmetry is the "minimal" one and, probably, the simplest model allowing for more or less adequate account of real nonergodic effects in large class of random ferromagnets and ferroelectric relaxors.

At last, the quadratic terms of the potential should reflect the competition between ferromagnetic (ferroelectric) and glassy orders. Thus we arrive at the unique form of potential up to forth order in $m_i$ obeying the above criteria,
\begin{eqnarray}
F\left( {\bf m} \right) = \frac{{\tau _f }}{{2N_0 }}\left( {\sum\limits_{i = 1}^{N_0 } {m_i } } \right)^2  + \frac{{\tau _g }}{{4N_0 }}\sum\limits_{i,j = 1}^{N_0 } {(m_i  - m_j )^2 }
\nonumber
\\
  + \frac{a}{4}\sum\limits_{i = 1}^{N_0 } {m_i^4 }  + \frac{b}{{4N_0 }}\left( {\sum\limits_{i = 1}^{N_0 } {m_i^2 } } \right)^2  - h\sum\limits_{i = 1}^{N_0 } {m_i } 
\label{eq:1}
\end{eqnarray}
Here $h$ is external field conjugate to the net magnetization of a sample
\[
\bar m = N_0^{ - 1} \sum\limits_{i = 1}^{N_0 } {m_i } 
\]
Note that this relation is the consequence of the adopted principle of mode equivalence which includes the supposition of equal numbers of sites participating in each mode. $N_0 $ in the denominators of Eq. (1) ensure the equal order of different terms' contributions at large $N_0 \gg 1$. Note also the absence in (1) the term
\[
\frac{c}{{N_0 }}\sum\limits_{i = 1}^{N_0 } {m_i } \sum\limits_{j = 1}^{N_0 } {m_j^3 }
\]
allowed formally by the above criteria. It can be removed by the linear transform of $m_i$ resulting in rescaling of $h$ and $\tau_f$ and appearance of two forth-order terms with three- and four-mode interactions which must be dropped due to assumed smallness of such interactions. Thus the presence in $F$ of this term amounts just to the $h$ and $\tau_f$  rescaling and it can be omitted.

The coefficients $a$ and $b$ in $F$ are some constants specific for a given disorder realization, while $\tau_f$  and $\tau_g$ are linear decreasing functions of temperature $T$ changing their signs at temperatures $T_f $ and $T_g$ correspondingly also being disorder dependent. Yet in the ferroelectric and ferromagnetic solid solutions all potential parameters must be the self-averaging quantities. It means that they depend only on the impurities' concentrations in accordance with the experimental data showing no noticeable variations of the properties of different samples with the same composition. Here we should note that this is true only for the solid solutions with deeply frozen disorder. In the relaxors with annealing mediated ordering of components such as PSN the parameters of the potential will also depend on the degree of ordering achieved at the high-temperature annealing.

Generally, the crystal with negligible disorder should have $T_g \ll T_f$ as in this case we may have just the ordinary transition into homogeneous ferro-phase without any traces of disorder. So the growth of disorder will result in the increase of $T_g$ along with the lowering of $T_f$. 
  
The difference between $T_f $ and $T_g$ must be small to make the expansion of $F$ in small $m_i$ meaningful. According to Eq. (\ref{eq:1}) below $T_f $ ferromagnetic order may set in, while below $T_g $ glassy states described by the  $N_0  - 1$ - dimensional order parameter composed of the independent $m_i - m_j$ components can appear favoring the antiparallel $m_i$ orientations. What actually results from the competition of glass- and ferro-order depends on the relation between $T_f $ and $T_g$  values as well as the rates of $\tau_f$  and $\tau_g$  decreasing. Finding the evolution of  $F\left( {\bf m} \right)$ minima for different $\tau_f$ and $\tau_g$  relations we can get the possible variants of phase sequences in crystals with random interactions. In spite of huge amount of glassy minima that $F\left( {\bf m} \right)$ may have, to find all of them is rather simple task in the present model which can be fulfilled by semi-analytical methods. 

\section{Thermodynamics of competing glass- and ferro-states in zero field}
\label{sec:2}
Using the notations $\left[ {m^k } \right] = N_0^{ - 1} \sum\limits_{i = 1}^{N_0 } {m_i^k } 
$, we can represent the potential density $f({\bf m}) = F({\bf m})/N_0 $ in the simple form
\[
f\left( {\bf m} \right) = \frac{{\tau _g }}{2}\left[ {m^2 } \right] + \frac{{\tau _f  - \tau _g }}{2}\bar m^2  + \frac{a}{4}\left[ {m^4 } \right] + \frac{b}{4}\left[ {m^2 } \right]^2  - h\bar m
\]
Differentiating $f\left( {\bf m} \right)$ with respect to $m_i$ we get the equations of state
\begin{equation}
\left( {\tau _g  + b\left[ {m^2 } \right]} \right)m_i  + am_i^3  = h + \left( {\tau _g  - \tau _f } \right)\bar m
\label{eq:2}
\end{equation} 
The solutions of these equations are minima of $f\left( {\bf m} \right)$ describing possible (meta)stable states if they render the positive definiteness of the matrix 
\begin{eqnarray*}
G_{i,j}  \equiv \frac{{\partial ^2 f({\bf m})}}{{\partial m_i \partial m_j }} = \delta _{i,j} \left( {\tau _g + b\left[ {m^2 } \right] + 3am_i^2 } \right)
\\
 + N_0^{ - 1} \left( {\tau _f  - \tau _g  + 2bm_i m_j } \right).
\end{eqnarray*}
 
	Let us consider first the homogeneously magnetized ferro-state with the equal $m_i  = m_0$. From Eq. (\ref{eq:2}) we have 
\begin{equation}
\tau _f m_0  + \left( {a + b} \right)m_0^3  = h
\label{eq:3}
\end{equation}
Thus at $h = 0$ the ferro-state appears at $\tau_f  < 0$ if $a + b > 0$. It is stable for
\begin{equation}
\tau _g  > \frac{{3a + b}}{{a + b}}\tau _f 
\label{eq:4}
\end{equation}
Turning to the possible glassy solutions of Eq. (\ref{eq:2}) with unequal $m_i$, we note that all $m_i$ obey the same equation so such solutions exist if there are several real roots to Eq. (\ref{eq:2}). Being of third order with respect to $m_i$ it can have one or three real roots. In the last case $m_i$  can acquire only two of the root values with the largest modules: one positive, $m_ +   > 0$, and one negative, $m_ -  < 0$, as only these two can make the coefficient at $\delta_{i,j}$ in matrix $\hat G$ positive which is necessary for the stability. Thus every glassy state is defined by the number, $N_+$, of $m_i$ having $m_+$  values (or the number of $m_-$ ones, $N_- = N_0 - N_+$). Hence in the glassy states
\[
\begin{array}{l}
\left[ {m^k } \right] = n_ +  m_ + ^k  + n_ -  m_ - ^k ,
\qquad
\bar m = n_ +  m_ +   + n_ -  m_ - 
\\
n_ \pm   \equiv N_ \pm  /N_0,
\qquad
 n_ +   + n_ -   = 1 \\
 \end{array}
\]
and to find a glass state with a given $ n_ + $   ($n_ -   $) we need to obtain the corresponding roots, $m_ \pm $, of Eq. (2) which now becomes
\begin{equation}
\left( {\tau _g  + b\left[ {m^2 } \right]} \right)m_ \pm   + am_ \pm ^3  = h + \left( {\tau _g  - \tau _f } \right)\bar m
\label{eq:5}
\end{equation}
Thus the present model may generally have the quasi-continuous set of states (local minima) defined by the parameter $n_+$ which changes in the interval $0 < n_ +   < 1$ in the infinitesimal steps $ \pm N_0^{ - 1} $. Note that there are  $
\left( \begin{array}{l}
 N_0  \\ 
 N_ +   \\ 
 \end{array} \right)
$ states with the same  $N_ +   = n_ +  N_0$ which differ by the permutations of $m_i$ and have the identical thermodynamic parameters due to the adopted permutation symmetry.
 
Introducing the variable
\[
x =  - m_ -  /m_ +  
\]
we get from Eq. (5) the following equations 
\begin{eqnarray}
\left( {1 - \frac{{\tau _f }}{{\tau _g }}} \right)\left( {n_ +   - xn_ -  } \right)R\left( {x,n_ +  ,\beta } \right) =
\nonumber
\\
x\left( {x - 1} \right)R\left( {x,n_ +  ,\beta } \right)^3 + \frac{h}{a}\left( {\frac{a}{{ - \tau _g }}} \right)^{3/2} 
\label{eq:6}
\\
m_ +   = \sqrt { - \tau _g /a} R\left( {x,n_ +  ,\beta } \right)
\label{eq:7}
\\
R\left( {x,n_ +  ,\beta } \right) \equiv \left[ {1 - x + x^2  + \beta \left( {n_ +   + n_ -  x^2 } \right)} \right]^{ - 1/2}
\nonumber
\end{eqnarray} 
Here $\beta  \equiv b/a$.

Thus we need to solve only one Eq. (6) for $x = x\left( {n_ +  ,\tau _g ,\tau _f ,h} \right)$  to obtain using Eq. (7) and $x$ definition full description of the thermodynamic properties of a crystal in the metastable states with a given $n_+$. So we have for the net spontaneous moment in such states 
\begin{equation}
\bar m = \left( {n_ +   - xn_ -  } \right)\sqrt { - \tau _g /a} R\left( {x,n_ +  ,\beta } \right)
\label{eq:8}
\end{equation}
and for the Edwards-Anderson glass order parameter
\[
q = \left[ {m^2 } \right] - \bar m^2  = \left( { - \tau _g /a} \right)n_ +  n_ -  \left( {1 + x} \right)^2 R\left( {x,n_ +  ,\beta } \right)^2. 
\]	
Further we consider the simplest case $a > 0, b > 0$ so the glassy metastable states appear at $\tau_g<0$ .

Then we have for the susceptibility  $\chi  = \frac{{\partial \bar m}}{{\partial h}}$
\begin{widetext}
\begin{equation}
 \chi ^{ - 1}  = \tau _f  - \tau _g  - \tau _g R\left( {x,n_ +  ,\beta } \right)^2 \frac{{\left( {1 + x} \right)\left( {2x - 1} \right)\left( {2 - x} \right) + 2\beta \left[ {\left( {2x - 1} \right)n_ +   + x^2 \left( {2 - x} \right)n_ -  } \right]}}{{\left( {2x - 1} \right)n_ +   + \left( {2 - x} \right)n_ -   + 2\beta n_ +  n_ -  \left( {1 + x} \right)}}
\label{eq:9} 
 \end{equation}
\end{widetext}

The equilibrium value of the potential for a given state is
\[
f_{eq} \left( {n_ +  ,\tau _g ,\tau _f ,h} \right) = \left( {\tau _g q + \tau _f \bar m^2  - 3h\bar m} \right)/4
\]
Here we omit the term arising from the degeneracy of glassy states
\[
 - \frac{T}{N}S_{config}  =  - \frac{T}{N}\ln \left( \begin{array}{l}
 N_0  \\ 
 N_ +   \\ 
 \end{array} \right)
 \sim \frac{{N_0 }}{N} \sim N^{ - d_f /d} 
\]
as it vanishes at large $N$.
Then we get for a given metastable state the entropy
\[
\begin{array}{l}
S =  - \frac{{\partial f_{eq} \left( {n_ +  ,\tau _g ,\tau _f ,h} \right)}}{{\partial T}} =  - \frac{1}{2}\left( {\tau '_g q + \tau '_f \bar m^2 } \right),
\\
\tau '_{f,g}  \equiv \frac{{\partial \tau _{f,g} }}{{\partial T}},
\end{array}
\]
and the heat capacity 
\begin{widetext}
\begin{eqnarray}
 \frac{C}{{T_g \chi }} = \left( {\tau '_f  - \tau '_g } \right)^2 \bar m^2
+ 2\tau '_g \left( {\tau '_f  - \tau '_g } \right)\frac{{\bar mm_ +  \left[ {n_ +  \left( {2x - 1} \right) - n_ -  x\left( {2 - x} \right)} \right]}}{{n_ +  \left( {2x - 1} \right) + n_ -  \left( {2 - x} \right) + 2n_ +  n_ -  \beta \left( {1 + x} \right)}}\nonumber \\
 + (\tau'_g)^2  \frac{{m_ + ^2 \left[ {n_ +  \left( {2x - 1} \right) + n_ -  x^2 \left( {2 - x} \right)} \right] + n_ +  n_ -  \left( {1 + x} \right)a^{ - 1} \left( {\tau _f  - \tau _g } \right)}}{{n_ +  \left( {2x - 1} \right) + n_ -  \left( {2 - x} \right) + 2n_ +  n_ -  \beta \left( {1 + x} \right)}} 
\label{eq:10} 
\end{eqnarray}
\end{widetext} 

For $a>0$, $b>0$ the stability conditions for the solutions of Eq. (\ref{eq:6}) providing the positive definetness of matrix $\hat G$ are represented by the  inequalities
\begin{equation}
1/2<x<2,
\qquad 
 \chi>0
\label{eq:11}
\end{equation}
Note that the condition $\chi >0$ ensures the stability with respect to the ferromagnetic fluctuations while at the boundaries of the first inequality the glassy instability occurs. 

Turning to the case $h = 0$ we can see that simple solutions of the equation of state (6) exist for   ${\rm  }n_ +   \to 1,0$ and  ${\rm  }n_ +   = n_ - = 1/2$. In the first case Eq. (\ref{eq:6}) tends to the zero-field limit of Eq. (\ref{eq:3}) and we have 
\[
\begin{array}{l}
\bar m = \sqrt{\frac{{ - \tau _f}}{{a + b}}},{\rm   }\chi ^{- 1} = - 2 \tau_f,{\rm   }q = 0,{\rm   }f_{eq}  = \tau _f \bar m^2 /4,\\
S =  - \tau '_f \bar m^2 /2,{\rm   }C = \frac{{T_g (\tau'_g)^2 }}{{2 \left( a + b \right)}}\\
\end{array}
\]
and Eqs.(\ref{eq:11}) defines the sector on the ($\tau_f, \tau_g$)-plane in which nearly homogeneous states with   ${\rm  }n_ +   \to 1,0$ are stable
\begin{equation}
\frac{{3 + \beta }}{{1 + \beta }}\tau _f  < \tau _g  < \frac{{3 + 4\beta }}{{4\left( {1 + \beta } \right)}}\tau _f  < 0
\label{eq:12}
\end{equation}
Note that such states exist only in the part of the region Eq. (\ref{eq:4}) where the homogeneously magnetized ferro-state is stable. This just means that the set of metastable states to which the states with ${\rm  }n_ +   \to 1,0$  belong has more narrow region of existence.

For the state with fully disordered moments (${\rm  }n_ +   = n_ -   = 1/2$ ) we have $x = 1 $ from Eq. (\ref{eq:6}) at $h=0$ so
\[
\begin{array}{l}
 \bar m = 0,{\rm   }\chi ^{{\rm  - 1}}  = \tau _f  - \tau _g \frac{{3 + \beta }}{{1 + \beta }},{\rm   }q = \frac{{ - \tau _g }}{{a + b}},{\rm   }f_{eq}  = \tau _g q{\rm /4}{\rm ,} \\ 
 {\rm   }S =  - \tau '_g q{\rm /2},{\rm     }C = \frac{{T_g (\tau'_g)^2 }}{{2\left( {a + b} \right)}} \\ 
 \end{array}
\]
and  $\chi  > 0$ requires that  
\begin{equation}
\tau _g  < \frac{{1 + \beta }}{{3 + \beta }}\tau _f 
\label{eq:13}
\end{equation}
along with $\tau _g  < 0$. It can be shown that the sector of stability of fully disordered states is the largest one and the stability regions of all other states belong to it, the most narrow sector being that of nearly homogeneous states (\ref{eq:12}). 

Further we find
\[
 \frac{{\partial f_{eq} }}{{\partial n_ +  }} =  - \frac{a}{4}\left( {m_ +   + m_ -  } \right)\left( {m_ +   - m_ -  } \right)^3 ,{\rm    }  
\]
Thus   $f_{eq} \left( {n_ +  ,\tau _g ,\tau _f ,h} \right)$ as function of $n_+$ has one extremum at 
\[
m_ +   + m_ -   = 0
\qquad
(x = 1),
\]
and we have at this point
\[
\left. {\frac{{\partial ^2 f_{eq} }}{{\partial n_ + ^2 }}} \right|_{x = 1}  = 8am_ + ^4 \chi \left( {\tau _f  - \tau _g } \right)\frac{{1 + \beta }}{{1 + 4\beta n_ +  n_ -  }}.
\]
So it is minimum at $\tau _g  < \tau _f $ and maximum at $\tau _g  > \tau _f $. Thus the states with ${\rm  }n_ +   = n_ -   = 1/2$ have the lowest potential at $\tau _g  < \tau _f $  while the global minimum is at the states with  ${\rm  }n_ +   \to 1,0$ at $\tau _g  > \tau _f $. 

\begin{figure*}
\centering
\includegraphics{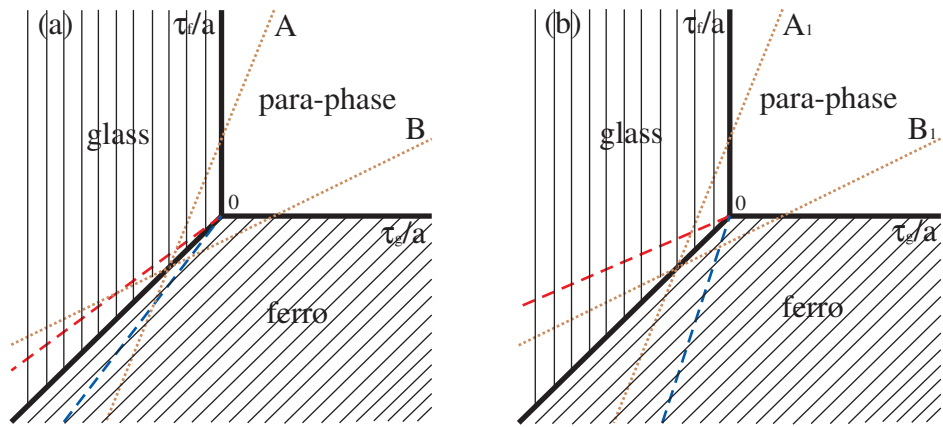}
\caption{\label{Fig.1} (Color online) Phase diagram of the model for $\beta = 5$ (a) and $\beta = 0.1$ (b). Thick lines denote the phase transitions between the phases. Upper dashed lines show the stability boundaries for the ferro-state (Eq. \ref{eq:4}) and lower dashed lines show that of  fully disordered states with $n_+ = 0.5$ (Eq. \ref{eq:13}). Dotted lines correspond to paths $\tau _f = 2.5\tau_g + 1.5a$ (A and $A_1$) and $\tau_f = 0.5(\tau_g - a)$ (B and $B_1$). }
\end{figure*}

Hence the first-order transition from the fully disordered glass to ferro-state takes place at $\tau _g  = \tau _f $. One can see from the above equations that potentials of these states indeed become equal at $\tau _g  = \tau _f $.  The phase diagram of the model is shown in  Fig. \ref{Fig.1}. The topology of the phase stability regions in it is the same that is found in the various microscopic models of disorderd magnets, cf. Refs. [\onlinecite{13}], [\onlinecite{30}-\onlinecite{32}].

In a specific crystal there is some linear relation between $\tau _f $ and $\tau _g $ both being linear functions of $T$. It defines a straight line on ($\tau_f, \tau_g$) plane which crystal follows under $T$ variations and, hence, the phase sequence proper for a given crystal. For example, if the relation $\tau _f  = \tau _g  + \tau _0$ holds, $\tau _0 $  being some constant, the crystal undergoes just one second-order transition - into the ordinary ferro-phase if   $\tau _0 <0$  or in the glass phase if $\tau _0 >0$. In the first case there is another qualitative change of thermodynamic properties (not a phase transition) - the subsequent appearance of numerous metastable states in addition to ferro-state begins when crystal reaches the regions of their stability belonging to the sector (\ref{eq:13}). Yet in these cases the application of field causes no drastic changes in the temperature dependencies of $\bar m$, $\chi$ and $C$ apart from the usual differences in their FC and ZFC values.

Much more pronounced field effects occur when the path on ($\tau _g , \tau _f $) plane intersects the line $\tau _g  = \tau _f $. Then glass or ferro-phase can appear as intermediate one in some finite temperature interval neighboring para-phase and first-order transition between them takes place at $\tau _g  = \tau _f $. Yet there are also different temperature behavior when path crosses the region of coexistence of ferro- and glass states (the sector in Fig. 1 bounded by the dashed lines) and when it always stays in this region. These four principally different paths are shown in Fig. 1. Dotted lines in it correspond to paths $\tau _f  = 2.5\tau _g  + 1.5a$ (A and $A_1$) and  $\tau _f  = 0.5(\tau _g - a)$ (B and $B_1$) which we use further to present some typical temperature variations of thermodynamic variables which differ essentially on them.

The spin-glass phase on paths B and $B_1$ is often called "reentrant" as crystal returns again on cooling to the equilibrium state with zero net magnetization. The possible microscopic mechanism of such reentrance is considered in Ref. [\onlinecite{9}].

We must note that here the first-order transition does not mean that at $\tau _g  = \tau _f $ the crystal do jump from one state to another on the laboratory time scale. Actually it must traverse the rough potential landscape and to overcome on the way a number of macroscopic barriers between the local minima (partially ordered states) to arrive to the global minimum. It takes an infinite time and the standard thermodynamics indeed predicts formally what will happen after infinite time when system will be fully relaxed. But actually one has just finite laboratory time to measure the thermodynamic quantities. 

It is quite usual situations in all nonergodic systems (i. e. those with metastable states) - they all have infinite (Arrhenius) relaxation time for the decay of metastable states into stable one irrespective of their number - to traverse just one macroscopic barrier also needs an infinite time the same as to traverse many of them. This circumstance causes no difficulties in the application of the results of the equilibrium thermodynamics to the known nonergodic systems such as ordinary ferroelectric and ferromagnets in their ferro-phases. The simple example is the description of hysteresis loop. Here the infinite-time thermodynamic prediction that spontaneous magnetization should reverse its direction just when the field change its sign is easily modified with due account of the long-living metastable states (in which field and magnetization have opposite signs). As they can not relax to the global minimum at the laboratory time and persist until they become unstable (at the coercive fields) one just include them in the description of the response to sufficiently slow (quasi-static) field thus obtaining upper and lower branches of hysteresis loop. Apparently, to apply the results of static thermodynamic to the dynamic process of field variation we must be sure that during it the system stays close to a local minimum. Hence, in this case the quasi-static condition means that the characteristic time of field variations is greater than the magnetization relaxation time in a local minimum.
	
In the same way we can get the description of quasi-static responses from the thermodynamic results for the present model. To do this one must assume that quasi-static regimes can be attained experimentally, i. e. that sufficiently slow temperature or field variations can be achieved on the laboratory time scale which ensure the location of system near some minimum during these variations. Then we just should take into account that once the system entered the metastable state it can not leave it in such quasi-static processes until that state becomes unstable.
	
Hence, the validity of this picture of the temperature evolution in which crystal is trapped in a local minimum and has the obtained above thermodynamic parameters crucially depends on the rate of cooling or heating. It should be sufficiently small to provide the quasi-static evolution of the system, that is the characteristic time of temperature variations, $t_0  = \left( {\frac{{d\ln T}}{{dt}}} \right)^{ - 1} $, must be greater than the largest magnetic relaxation time in the specific minimum. The same is true for characteristic time of field variations. One may expect that when we are not close to $T_g$ nearly homogeneously magnetized states with $\bar m \sim m_0 $ would have rather moderate relaxation times comparable to those the crystal has deep in para-phase. Also far from $T_g$ the magnetic susceptibility of highly disordered states with  $\bar m \ll m_0 $ may be very small making the amplitude of magnetization relaxation in them small too. Then we would observe their nearly static parameters even for $t_0$ less than relaxation times which can spread up to huge macroscopic values in such states. If  $\chi$ reaches large values close to $T_g$  in the states with low $\bar m \ll m_0 $ their thermodynamic description would become useless due to the strong aging effects in them. Then one should necessarily resort to the dynamic theory to get the adequate account of their (dynamic) properties. 

Provided the above quasi-static conditons are fullfilled another question inherent to such highly nonergodic systems arises - to which of numerous stable states the system will go after its present state becomes unstable?  Strictly speaking, one should again turn to the dynamics to answer it, yet, as we will see, there is actually no vast choice of possibilities in the framework of the heuristic assumption that system adopts the most smooth variations of its average magnetization. 

Thus crystal can stay in metastable state for a very large (Arrhenius) time and leave it only when it becomes unstable. Before this the only way to bring the nonergodic crystal to some other state is to apply the external field which could make the present state unstable. This circumstance made former investigators to believe that ferroelectric relaxors have no phase transitions exhibiting only "field - induced" ferroelectricity \cite{17}.
	  
\begin{figure*}
\centering
\includegraphics{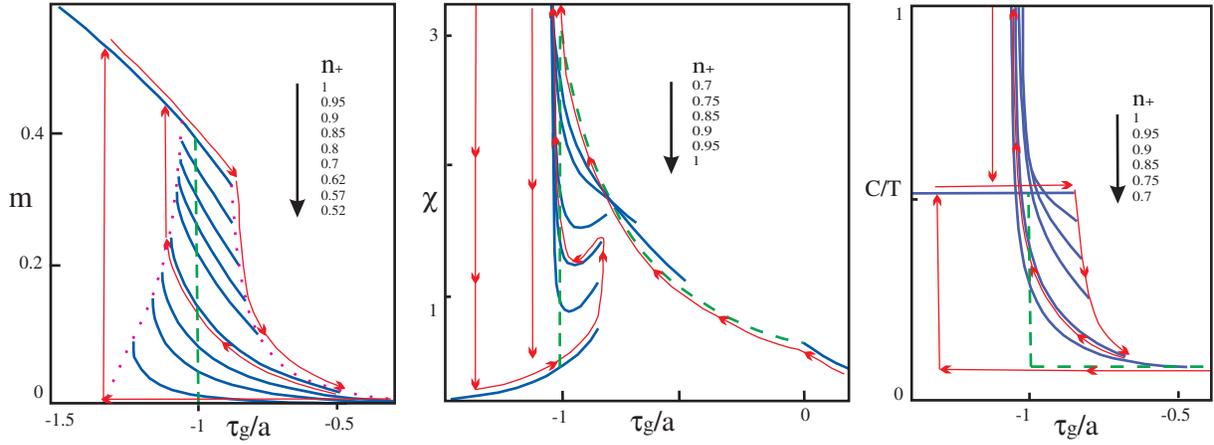}
\caption{\label{Fig.2} (Color online) Temperature dependencies of metastable states' magnetization, susceptibility and heat capacity in zero field for $\beta = 5$ corresponding to path A in Fig.\ref{Fig.1}(a). Dotted lines show the stability boundaries and dashed ones represent the parameter's variations predicted by the equilibrium thermodynamics (those of the states with $n_+ =1/2$ in the present case). Directed lines show the evolution of magnetization in real-time quasi-static regime.}
\end{figure*}

\begin{figure*}
\centering
\includegraphics{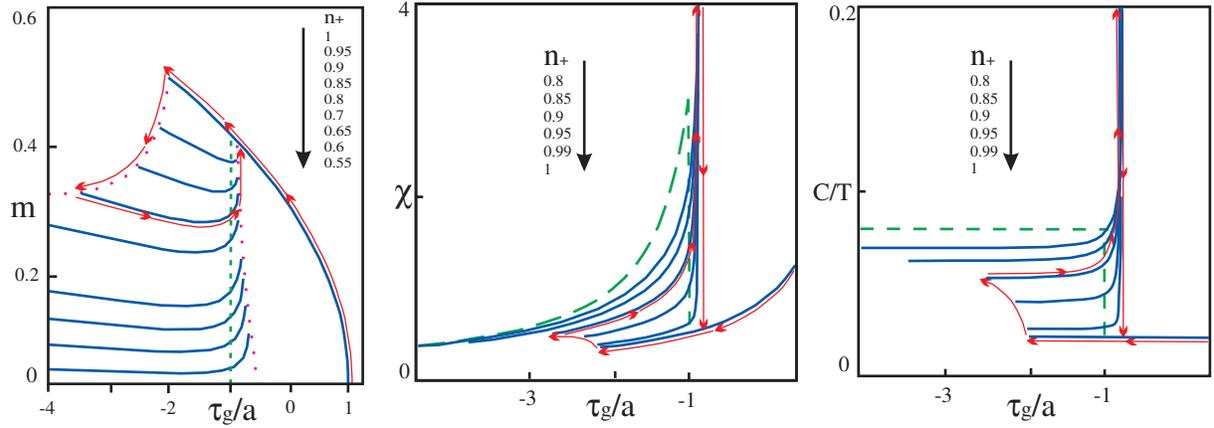}
\caption{\label{Fig.3} (Color online) The same as in Fig. \ref{Fig.2} for path B in Fig. \ref{Fig.1}(a).}
\end{figure*}

\begin{figure*}
\centering
\includegraphics{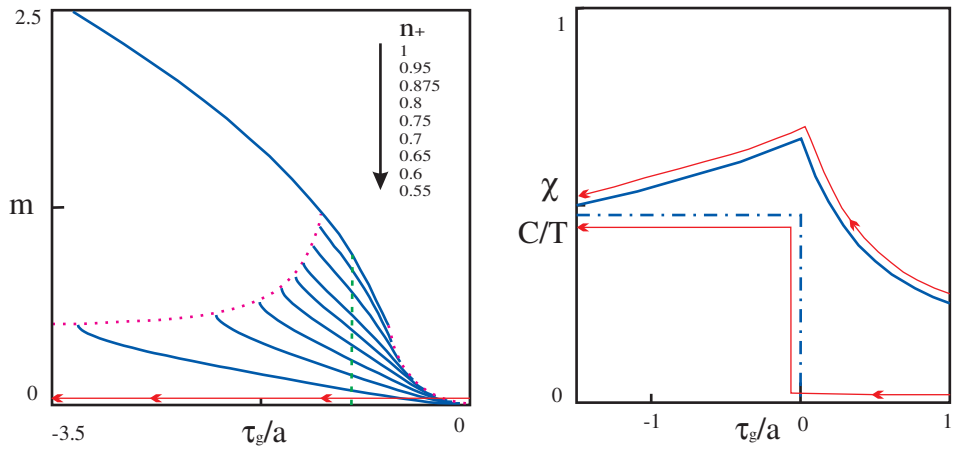}
\caption{\label{Fig.4} (Color online) The same as in Fig. \ref{Fig.2} for path $A_1$ in Fig.\ref{Fig.1}(b). Dash-dotted line corresponds to $C/T$. $\chi$ and $C$ are shown only for the $n_+=1/2$ state as others are inaccessible at $h=0$}
\end{figure*}

\begin{figure}
\includegraphics{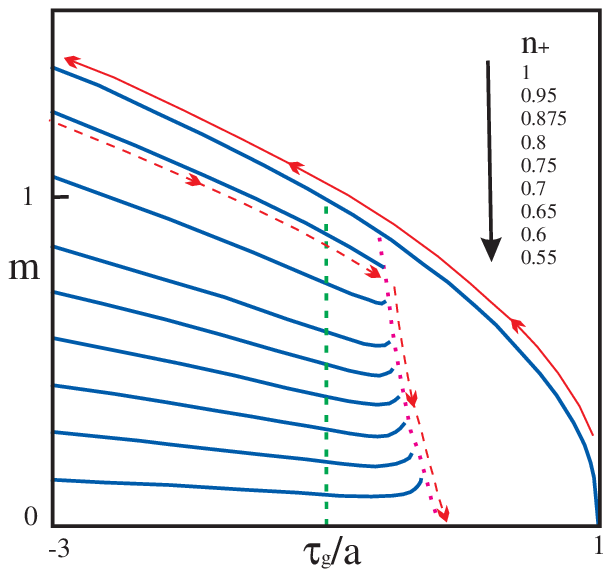}
\caption{\label{Fig.5} (Color online) Temperature dependencies of metastable states' magnetization in zero field for $\beta = 0.1$ corresponding to path $B_1$ in Fig. \ref{Fig.1}(b). Dashed directed line show the decay of nearly homogeneous state's magnetization in the infinitesimal field. The other lines have the same meaning as in Fig.\ref{Fig.2}.}
\end{figure}

All these considerations must be taken into account in the interpretation of the temperature dependences of metastable parameters described by the above equations and shown in Figs. \ref{Fig.2} - \ref{Fig.5}. Note that to plot them one does not need to solve the equation of state (\ref{eq:6}) as it provides along with the expression for a thermodynamic variable the parametric representation of its temperature dependence, $x$ being the parameter varying in the region of stability of a given state.

The temperature dependencies of metastable states' magnetizations in  Figs. \ref{Fig.2} - \ref{Fig.5} are bounded by the lines at which states become unstable. In the glass phase they correspond to $x=1/2, 2$ so their equations follow from Eqs. \ref{eq:6}, \ref{eq:8} at these values of $x$
\begin{eqnarray}
\bar m^2  = \left( {1 + \frac{3}{{2\beta }} \pm \sqrt {\left( {1 + \frac{3}{{2\beta }}} \right)^2  + \frac{{2\tau _g }}{{\beta \left( {\tau _f  - \tau _g } \right)}}} } \right)^3
\nonumber\\
\times\frac{{\tau _f  - \tau _g }}{{2a}}. 
\label{eq:14}
\end{eqnarray}
Here minus and plus signs are for (Figs. \ref{Fig.2}, \ref{Fig.4}) and (Fig. \ref{Fig.3}) correspondingly. In the ferro-phase this relation is not generally valid as some states seize to exist due to the ferromagnetic instability ($\chi \to \infty$). In this case it is hard to get the compact analog of Eq. (\ref{eq:14})

As crystal leaves its state only when it becomes unstable, it is rather evident that the quasi-static heating (Fig. \ref{Fig.2}) or cooling (Fig. \ref{Fig.3}) of crystal being in the ferro-state beyond its stability point in the glass phase makes it to go through the succession of metastable states on the boundary of their stability regions. So the self-organized criticality shows up in these processes as system is permanently unstable in finite temperature intervals and its evolution proceeds via small but macroscopic avalanches of spin upturns.

Eqs. (\ref{eq:14}) describe the temperature dependencies of magnetization in these regimes. In particular, for the vanishing of magnetization on heating in Figs. \ref{Fig.2}, \ref{Fig.4} we get at $\tau_g \to 0$
\[
\bar m = \frac{{2a}}{{\left| {\tau _f } \right|}}\left( {\frac{{ - \tau _g }}{{3a + 2b}}} \right)^{3/2}. 
\]

Also the specific temperature hysteresis occurs in the cooling-heating-cooling cycles (Fig. \ref{Fig.2}) and cooling-heating ones (Fig. \ref{Fig.3}) starting on the paths A and B from the high-temperature phases (in the presence of infinitesimal positive field to have $\bar m>0$). These cycles are shown by the directed lines on the temperature dependencies of thermodynamic parameters. Their remarkable property is that they allow to enter a variety of metastable states (actually all in the range $0.5 < n_+ < 1$ in the case of intermediate glass phase in Fig.  \ref{Fig.2} and those with $0.628  < n_+ < 1$ having the limited stability regions in Fig.  \ref{Fig.3}). For example, in the case of Fig. \ref{Fig.2} we can stop heating at some $T$ when crystal traverses the stablility boundaries of metastable states and start cooling it thus trapping crystal in the state it was in at the stop.

Note the sharp rise of the susceptibilities and heat capacities of metastable states in Figs. \ref{Fig.2}, \ref{Fig.3}. Actually this denotes the divergence of $\chi$ and $C$ (cf. Eq.(\ref{eq:10})) due to the ferromagnetic instability metastable states experience in the ferro-phase at $\tau_f > \tau_g$ ($\tau_g/a < -1$  in Fig. \ref{Fig.2} and $\tau_g/a > -1$  in Fig. \ref{Fig.3}). 

On the path $A_1$ the appearance of intermediate glass phase in zero field is marked by the jumps in heat capacity and $\chi$ temperature derivative (Fig. \ref{Fig.4}) while on the path $B_1$ the intermediate ferro-phase (Fig. \ref{Fig.5}) has usual Curie - Weiss anomaly of $\chi$  and the usual jump in heat capacity (not shown). In these cases the appearance of metastable states can not be observed in zero-field experiments - at all $T$ the crystal will stay in the state it enters first on cooling. But the presence of metastable states can be easily revealed in the glass phase. Here stopping at some $T$ and applying for some time the field (above some threshold one needed to make the former state unstable) one can end up in every metastable state choosing the value of the applied field. In the further temperature variations the crystal will stay in the so prepared state until it vanishes.

\section{Thermodynamics in finite field}
\label{sec:3}
\begin{figure*}
\centering
\includegraphics{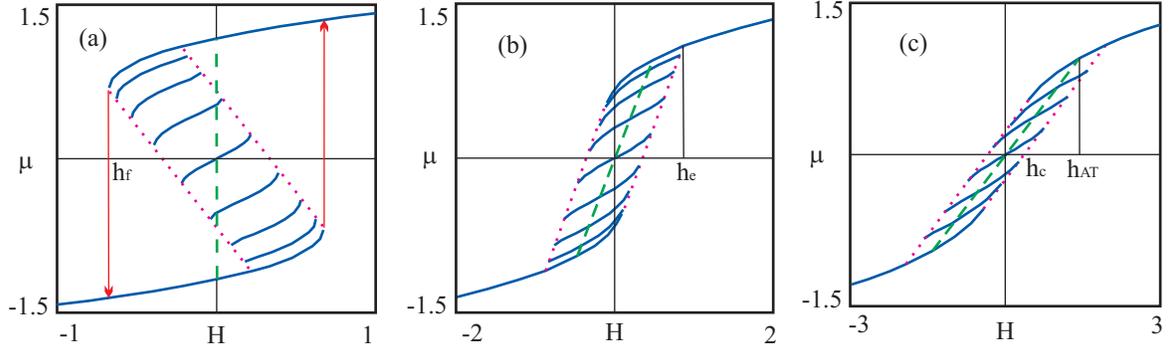}
\caption{\label{Fig.6} (Color online) Field dependencies of metastable states' magnetizations for $\beta  =  0.1$ on the path $A_1$, $\mu  \equiv \bar m\sqrt{-a/\tau _g} $,   $H \equiv h\sqrt {-a/\tau _g^3 }$, (a) - $\tau_g=-1.1a$, (b) - $\tau_g=-0.75a$, (c) - $\tau_g=-0.5a$}.
\end{figure*}

\begin{figure*}
\centering
\includegraphics{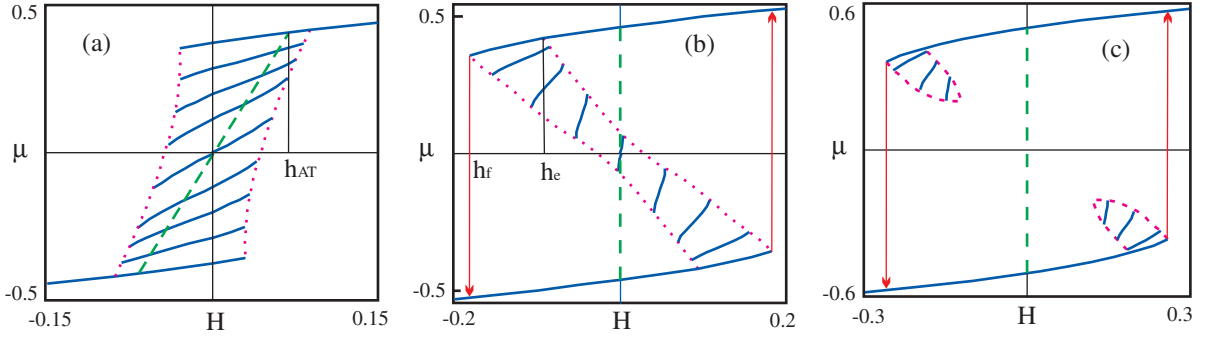}
\caption{\label{Fig.7} (Color online) The same as in Fig. 6 for $\beta = 5$ and path B. (a) - $\tau_g=-1.5a$, (b) - $\tau_g=-0.65a$, (c) - $\tau_g=-0.5a$}
\end{figure*}

Further we examine the properties of metastable states in finite fields. In the glass phase at $\tau _g  < \tau _f $  all metastable states always have $\chi >0$ so they have not ferromagnetic instabilities here. Again, substituting $x = 1/2$ and $x = 2$ in Eqs. (\ref{eq:6}, \ref{eq:9}) we get the parametric representation of two lines on ($\bar m, h$)-plane bounding the region where these states exist ($0 < n_+ < 1$ being the parameter). Excluding $n_+$ we get the equation two real roots of which define the couple of these lines
\begin{eqnarray}
3a + b\left[ {2 + \bar m\left( {\frac{{2a}}{{h + \left( {\tau _g  - \tau _f } \right)\bar m}}} \right)^{1/3} } \right]\nonumber\\
 + \tau _g \left| {\frac{{2a}}{{h + \left( {\tau _g  - \tau _f } \right)\bar m}}} \right|^{2/3}  = 0\label{eq:15}
\end{eqnarray}
These roots have   $h + \left( {\tau _g  - \tau _f } \right)\bar m > 0$ and   $h + \left( {\tau _g  - \tau _f } \right)\bar m < 0$ with the interval $0 < n_+ < 1$ being swept when $h$ varies correspondingly in the intervals $ - h_f  < h < h_e$ ,$- h_e  < h < h_f$
\begin{eqnarray}
h_f  = \sqrt {\frac{{ - \tau _g }}{{3a + b}}} \left( {\tau _f  - \tau _g \frac{{1 + \beta }}{{3 + \beta }}} \right),\label{eq:16}\\
h_e  = \sqrt {\frac{{ - \tau _g }}{{3a + 4b}}} \left( {\tau _f  - 4\tau _g \frac{{1 + \beta }}{{3 + 4\beta }}} \right)\label{eq:17}
\end{eqnarray}
The magnetization curves $\bar m= \bar m(h,n_+)$ of metastable states defined by Eqs. (\ref{eq:6}, \ref{eq:8}) fill the region bounded by the line segments of Eq. (\ref{eq:12}) as Figs. \ref{Fig.6}, \ref{Fig.7} show. Again the self-organized criticality appears in these segments as system is permanently unstable in them and its evolution proceeds via the avalanches of spin upturns.
Such avalanches and small magnetization jumps are shown to exist in hysteresis loops of Sherrington-Kirkpatrick model \cite{11}, \cite{12}.
 
At $\left| h \right| > h_e $ all metastable states vanish leaving the only stable homogeneous state with magnetization curve $m_0 \left( h \right)$ given by Eq. (\ref{eq:3}). As it was discussed above the crystal will stay in a local minimum until it vanishes so Fig. \ref{Fig.6} represents possible forms of hysteresis loops for quasi-static periodic field. If amplitude of such field $\left| h \right| > h_e $ the form of quasi-static loop is completely described by Eqs. (\ref{eq:3}, \ref{eq:12}). For smaller amplitudes the inner loops inside the outer one will be observed formed by the boundaries of Eq. (\ref{eq:12}) and $\bar m = \bar m\left( {h,n_ +  } \right)$ and $\bar m = \bar m\left( {h,1-n_ +  } \right)$ curves with some $n_+$ defined by the field amplitude. 

We can find the analog of the coercive field $h_c$ for the glassy loops putting $\bar m=0$  in Eq. (\ref{eq:15}),
\[
h_c  = 2a\left( {\frac{{\left| {\tau _g } \right|}}{{3a + 2b}}} \right)^{3/2} 
\]

The inclined hysteresis loop, see Figs. \ref{Fig.6}(b,c), \ref{Fig.7}(a), is the hallmark of glass phase; such loops are ubiquitous in random magnets and relaxors. They are found in the numerical simulation of spin glass phases of 2$d$ short-range random-bond Ising model \cite{10}, Sherrington-Kirkpatrick model \cite{11}, \cite{12}, random anisotropy model \cite{13} as well as in the local mean-field simulations \cite{7}, \cite{8}. Qualitatively the transition at $\tau_f=\tau_g$  manifest itself by the change of loop form, inclined in the glass phase it becomes the ordinary one with the vertical sides in the ferro-phase as disordered metastable states can not be reached in the quasi-static periodic field, see Fig. \ref{Fig.6}(a). It is shown that such change of hysteresis loop takes place in Sherrington-Kirkpatrick model with the growth of ferromagnetic exchange \cite{11}.

Also smaller loops vanish in the ferro-phase as the field evolution of all metastable states shown on Fig. \ref{Fig.6}(a), \ref{Fig.7}(b,c) ultimately ends by the transition into the homogeneous one. For $b = 0$ it happens strictly at  $\tau_f=\tau_g$ while for finite $b>0$ inclined loop transforms into rectangular one already in the glass phase slightly before the thermodynamic transition into the ferro-phase. Generally this change takes place when $h_c$ becomes less than $-h_f$, that is at 
\[
\tau _f  = \left( {\frac{{1 + \beta }}{{3 + \beta }} + 2\sqrt {\frac{{3 + \beta }}{{\left( {3 + 2\beta } \right)^3 }}} } \right)\tau _g 
\]
	
It is also easy to get for all $h$ the equilibrium magnetization $m_{eq}$   corresponding to the states with the lowest potential having $x=1$ in glass phase as was shown above. From Eqs. (\ref{eq:6}), (\ref{eq:8}) we have
\begin{equation}
\bar m_{eq}  = \frac{h}{{\tau _f  - \tau _g }}
\label{eq:18}
\end{equation}	
for the equilibrium states with
\begin{eqnarray*}
n_ + ^{eq}  = \frac{1}{2}\left( {1 + \frac{h}{{h_{AT} }}} \right)
\\
h_{AT}  = \frac{{\sqrt { - \tau _g } \left( {\tau _f  - \tau _g } \right)}}{{\sqrt {a + b} }}\\
\end{eqnarray*}	
The Almeida-Thouless field $h_{AT}$ marks the thermodynamic transition between the completely ordered equilibrium phase at   $\left| h \right| > h_{AT}$ to the sequence of disordered ones at $\left| h \right| < h_{AT}$.

For the disordered states $\bar m_{eq}$ is represented in the interiors of hysteresis loops in Figs. (\ref{Fig.6}, \ref{Fig.7}) by dashed lines with tangents defined by the equilibrium susceptibility
\begin{equation}
\chi _{eq}  = \frac{{\partial \bar m_{eq} }}{{\partial h}} = \frac{1}{{\tau _f  - \tau _g }}
\end{equation}
and ending at   $h =  \pm h_{AT} $.

Generally $\chi_{eq}$ is unobservable quantity as to follow the relation in Eq. (\ref{eq:18}) under the field variations the crystal has to overcome the macroscopic barriers in series of first-order phase transitions (stepping in them from the state with some $n_+$ to that with $n_ +   \pm N_0^{ - 1} $) which are not shown in Figs. \ref{Fig.6}, \ref{Fig.7} (see Fig.1 in Ref. [\onlinecite{26})]. As we discussed above, this would need a very long time and crystal will stay in the local minimum on the laboratory time scales. So the only susceptibility measurable under field variations is that given by Eq. (\ref{eq:9}) which defines the tangents of the magnetization curves $\bar m = \bar m\left( {h,n_ +  } \right)$ of the specific metastable state at a given field. It is always less than $\chi_{eq}$, thus in the equilibrium states with $x = 1$
\[
\chi ^{ - 1}  = \tau _f  - \tau _g \left[ {1 + \frac{{2h_{AT}^2 }}{{h_{AT}^2  + \beta \left( {h_{AT}^2  - h^2 } \right)}}} \right]
\]
 while at the right ($x =1/2$) and left ($x = 2$) boundaries of hysteresis loop in glass phase we have 
\[
\chi ^{ - 1}  = \tau _f  - \tau _g \left[ {1 + \frac{{6\beta }}{{\left( {3 + 4n_ \pm  \beta } \right)\left[ {3 + \beta \left( {1 + 3n_ \pm  } \right)} \right]}}} \right]
\]
with plus and minus sign correspondingly.

\begin{figure}
\includegraphics{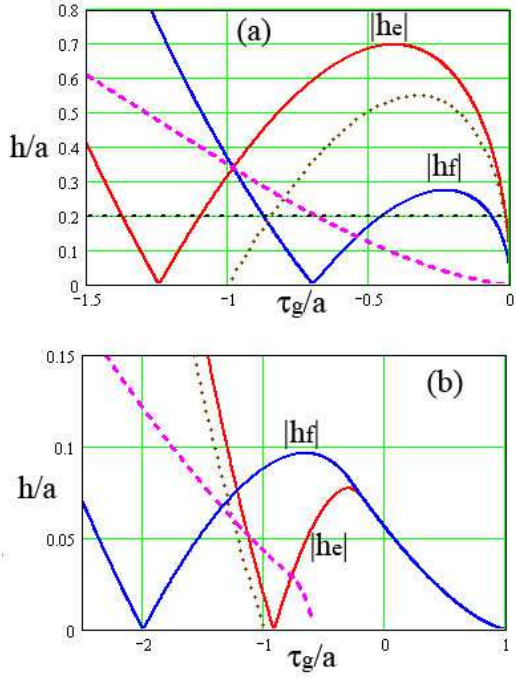}
\caption{\label{Fig.8} (Color online) Temperature dependencies of characteristic fields; (a) - $\beta = 0.1$ on the path $A_1$, dash-dotted line represents $h = 0.2a$, (b) -  $\beta = 5$ and path B. Dashed lines - $h_{AT}$, dotted ones - $h_c$. Solid lines show the modules of $h_f$ and $h_e$.}
\end{figure}

Yet in the case $b \ll a$ $\left( {\beta  \ll {\rm 1}} \right)$ $\chi_{eq}$ can be determined just from the shape of hysteresis loop. Indeed, Eq. (\ref{eq:15}) gives then for its boundaries the almost straight lines with tangent equal to $\chi_{eq}$, see Figs. \ref{Fig.6}(b, c). But for large $\beta$ this relation is lost owing to the intricate form of the loop as Figs. \ref{Fig.7}(a) show. 
For small $\beta$ the shape of hysteresis loop in the glass phase is mostly defined by the fields $h_e$ and $h_f$. In the ferro-phase ( $\tau _f  < \tau _g $) the loop boundaries are no longer given by Eq. (\ref{eq:15}) as here some states vanish due to ferromagnetic instability before reaching the glassy one at $x = 1/2$ or $x = 2$. Also in the ferro-phase the expression for the field $h_f$, Eq. (\ref{eq:16}) changes to 
\begin{equation}
h_f  =  - \frac{2}{3}\sqrt {\frac{{ - \tau _f^3 }}{{3\left( {a + b} \right)}}}
\label{eq:19} 
\end{equation}	
in the interval
\[
\frac{{3 + \beta }}{{3\left( {1 + \beta } \right)}}\tau _f  < \tau _g  < \frac{{3 + 4\beta }}{{12\left( {1 + \beta } \right)}}\tau _f .
\]
The field in Eq. (\ref{eq:19}) is just the coercive field for the homogeneous ferro-state. Note that at   
\[
\tau _g  = \frac{{3 + 4\beta }}{{12\left( {1 + \beta } \right)}}\tau _f 
\]	
the field $h_f$ and $h_e$ merge at this value so at   
\[
\tau _g  > \frac{{3 + 4\beta }}{{12\left( {1 + \beta } \right)}}\tau _f 
\]
disordered metastable states no longer exist at all $h$ and we have the ordinary ferromagnetic loop. The temperature dependencies of the loop's characteristic fields for paths $A_1$ and B are shown in Fig. \ref{Fig.8}, in the ferro-phase $h_c$  and $h_e$ were found numerically.  

\begin{figure*}
\centering
\includegraphics{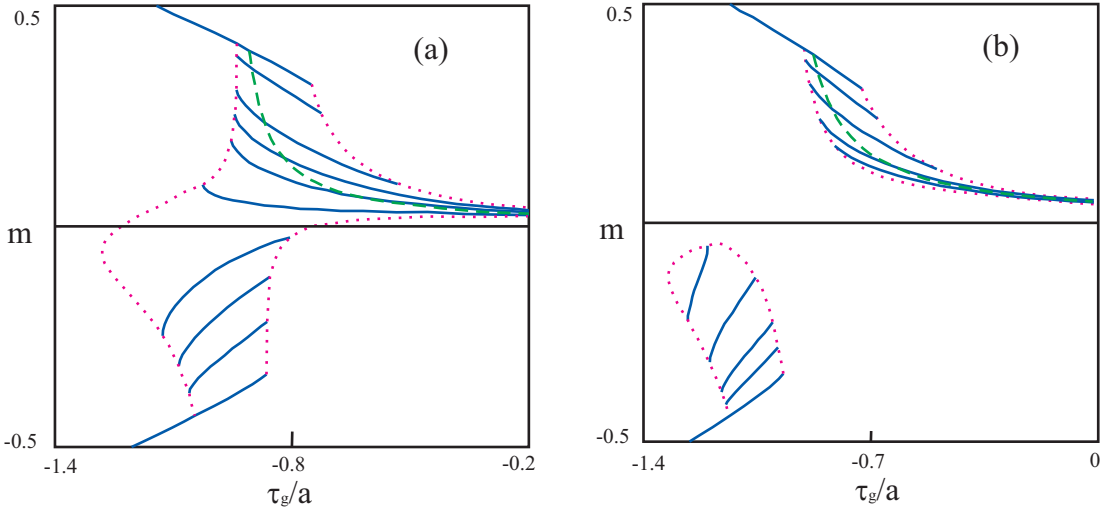}
\caption{\label{Fig.9} (Color online) Temperature dependencies of metastable states' magnetizations for $\beta = 5$ corresponding to the path A in the fields $h = 0.03a$ - (a) and $h = 0.07a$ - (b).}
\end{figure*}

\begin{figure*}
\centering
\includegraphics{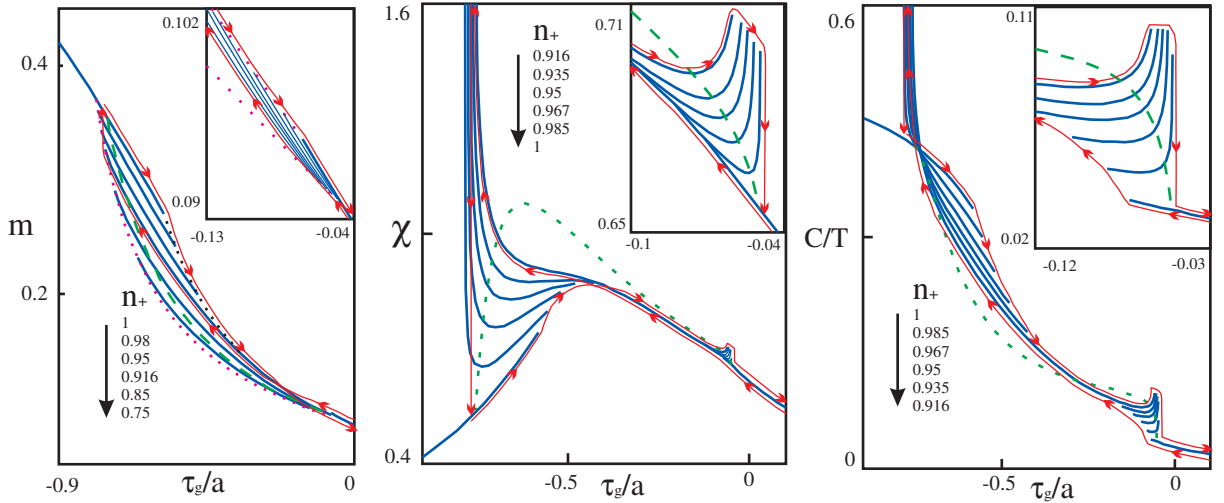}
\caption{\label{Fig.10} (Color online) Temperature dependencies of metastable states' magnetizations, susceptibility and heat capacity in a field $h = 0.13a$ for $\beta  = 5$ corresponding to the path A. Insets show the high-temperature anomalies.} 
\end{figure*}

\begin{figure*}
\centering
\includegraphics{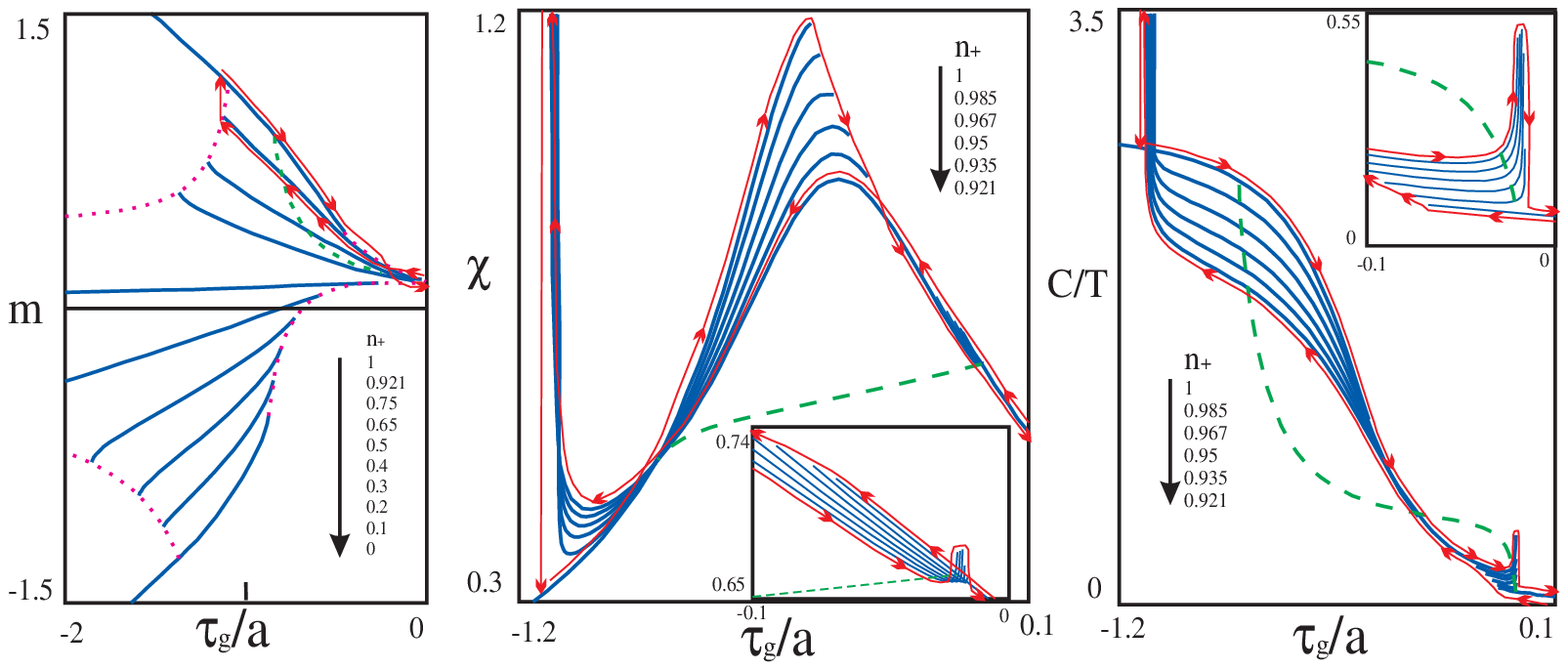}
\caption{\label{Fig.11} (Color online) Temperature dependencies of metastable states' magnetizations, susceptibility and heat capacity in a field $h = 0.2a$ for $\beta  =  0.1$ corresponding to the path $A_1$. Insets show the high-temperature anomalies. }
\end{figure*}

Now we can turn to the temperature dependencies of metastable states' parameteres in a constant field. In general these states gradually vanish at larger fields, first the most disordered ones, as one may expect. Fig. \ref{Fig.9} illustrates this process for the path A. 
Temperature dependencies of $\bar m$, $\chi$ and $C$ in constant field are closely related to that of $h_f$ and $h_e$. Thus for paths A and $A_1$ $h_f$  first grows on cooling and then diminishes after reaching the maximum, see Fig. \ref{Fig.8}(a) for path $A_1$ where $h_{f,max}\approx 0.28a $ is attained at $\tau _g  \approx  - 0.24a$. Also $h_e >h_f$ exhibits the similar behavior. This results in the essential changes in $\chi$ and $C$ dependencies for field-cooling and field-heating in Fig. \ref{Fig.10} and, especially, in Fig. \ref{Fig.11} as compared to those in Figs. \ref{Fig.2}, \ref{Fig.4} for $h= 0$.
In the insets of Figs. \ref{Fig.10}, \ref{Fig.11} there are slight anomalies of these parameters when crystal enters on cooling the interval where $h<h_f$ and leaves the homogeneous state to join the sequence of glassy ones with $n_+ <1$ until it reaches the first state being stable throughout all this interval. It varies for different paths and fields. In Figs. \ref{Fig.10}, \ref{Fig.11} they are those with $n_+ = 0.916$ and $n_+ = 0.921$ correspondingly. This process is nothing else than the descend throughout the upper branch of the growing hysteresis loop. In further cooling the crystal stays in this state until it vanishes due to ferromagnetic instability near the point where $h \approx h_e $ (at another branch of the loop having now the rectangular form, see Fig. \ref{Fig.6}(a)) for path $A_1$). So $\chi$ and $C$ diverge at this point according to Curie-Weiss law and crystal returns to the ferro-state. This is new feature for path $A_1$ which is absent in zero field, Fig. \ref{Fig.4}. On heating from this state $n_+$ again starts diminishing at the lower boundary of the interval where $h<h_f$  reaching $n_+ =0.916 (0.921)$ (upper branch of the loop). Further heating in this state ends up at the upper boundary of  $h<h_f$ interval in the sequence of states with $n_+ $ growing up to 1 (lower branch of the loop). This cause the slight high-temperature spikes in $\chi$ and $C$ manifesting the glass instabilities of these states.

Here we may note that Fig. \ref{Fig.11} explains the origin of additional pronounced peak in $\chi$ appearing in relaxor PMN in FC regime \cite{18,19}. It is the consequence of the instability of nearly homogeneous state the crystal enters on field-cooling. Moreover, the behavior of PMN dielectric susceptibility observed in all field-cooling-field-heating cycle \cite{18,19} resembles qualitatively that of $\chi$ in Fig. \ref{Fig.11}. The appearance of additional spikes of magnetic susceptibility in FC regime was also registered in random ferromagnets $PrNi_{0.3}Co_{0.7}O_3$ \cite{20} and $(Fe_{0.17}Ni_{0.83})_{75}P_{16}B_6Al_3$ \cite{21}.

We should also note that magnetizations in the regimes of cooling from the para-phase shown in Figs. \ref{Fig.10}(a),  \ref{Fig.11}(a) are the field-cooled ones, $m_{FC}$. These figures represent also the zero-field-cooled magnetizations,  $m_{ZFC}$, obtained at turning on the field after cooling the sample in zero field. They are the metastable curves with $n_+ = 1/2$ (if they exist) or the stability boundaries corresponding to the lower branch of inclined hysteresis loop or to the homogeneous states lines when the loop is rectangular as in Figs. \ref{Fig.10}(a). Indeed, in zero-field cooling process on paths $A$ and $A_1$  the system is trapped in the $n_+ = 1/2$ state, see Figs. \ref{Fig.2}(a), \ref{Fig.4}(a), and stays in it after turning on the field or goes to the lower loop branch or to the $n_+ = 1$ state depending on the loop form and field strength. So in Fig. \ref{Fig.10}(a) $m_{ZFC}$ corresponds to the lower stability line while in Fig. \ref{Fig.11}(a) it follows $n_+ = 1/2$ curve and then jump to $n_+ = 1$ curve when former seized to exist. 

\begin{figure*}
\centering
\includegraphics{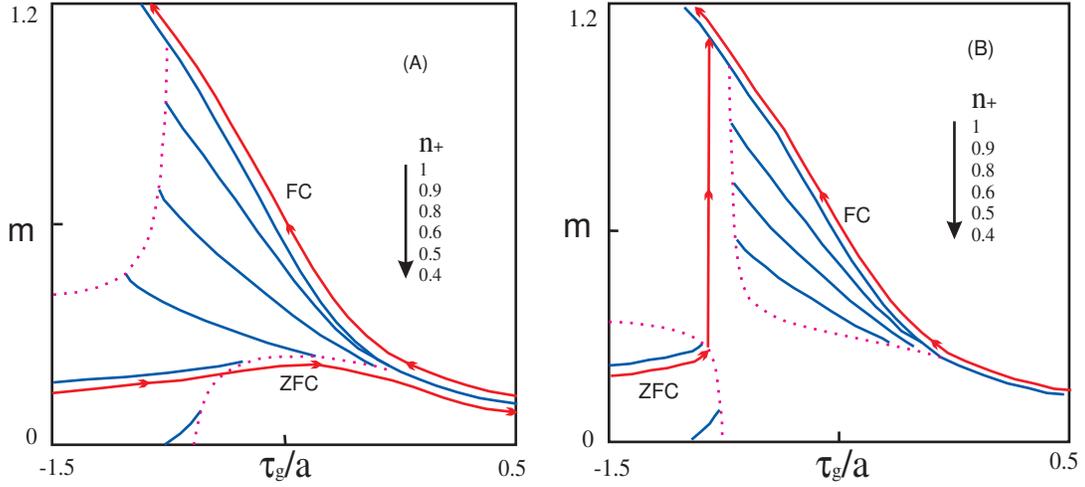}
\caption{\label{Fig.12} (Color online) Temperature dependencies of $m_{ZFC}$ and $m_{FC}$ on path $A_1$ in field $h = 0.3a$ (A)and $h = 0.35a$ (B).}
\end{figure*}

Here we should note that often in experiments and simulations another definition of $m_{ZFC}$ is used, namely, the magnetization registered in field-heating after zero-field-cooling process.  As Fig. \ref{Fig.10}(a) shows this is the ill-defined quantity as it will depend in this case on the temperature at which the turn from cooling to heating takes place. Yet when the field at the turning point leaves the $n_+=1/2$ state stable as for path $A_1$ in Fig. \ref{Fig.11}(a) the field-heating after zero-field-cooling protocol gives a unique $m_{ZFC}$ the same as in simple ZFC process described above. The detailed picture of $m_{ZFC}$ and $m_{FC}$  behavior for path $A_1$ is shown in Fig.\ref{Fig.12}. The qualitativly similar temperature dependences of these magnetizations are found in local mean-field simulations \cite{7} and in field-heating after zero-field-cooling experiments in PMN \cite{18}.

\begin{figure*}
\centering
\includegraphics{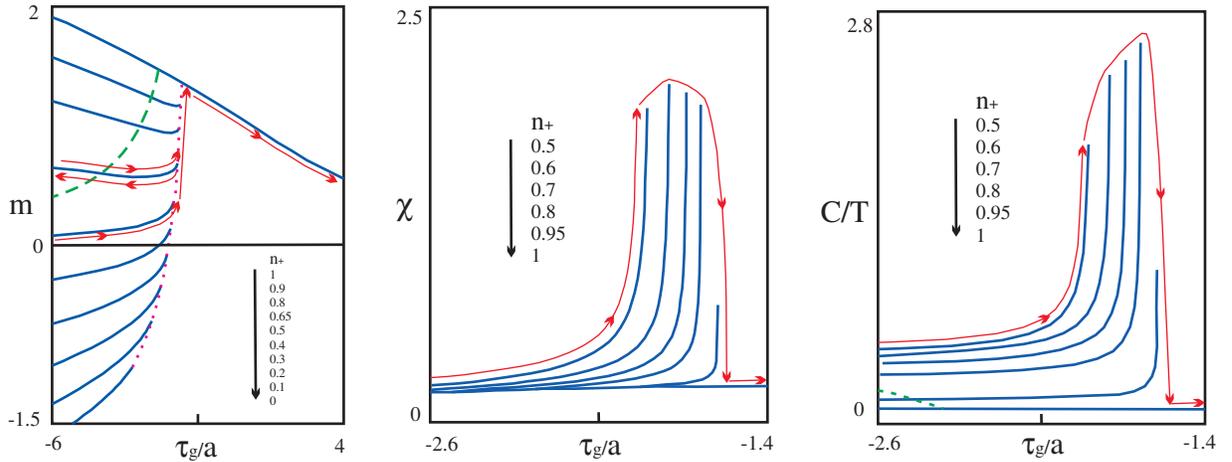}
\caption{\label{Fig.13} (Color online) Temperature dependencies of metastable states' magnetizations, susceptibility and heat capacity in a field $h = a$ for $\beta  =  0.1$ corresponding to the path B.}
\end{figure*}

The applied field cause also some peculiarities in temperature dependencies of $\bar m$, $\chi$ and $C$ when ferro-phase is intermediate between para- and glassy ones (path B). As Fig. \ref{Fig.13} shows they appear in field-heating regime of crystal being initially in the most disordered state with $n_+ = 1/2$. After reaching the limit of stability of this state the crystal enters the sequence of states with larger $n_+$ on the boundaries of their existence regions until it joins the fully homogeneous ferro-state with $n_+ = 1$. This process is accompanied by the spikes in $\chi$ and $C$. Along with it the reverse process of the demagnetization of almost homogeneous initial state on heating is possible in the infinitesimal field. It is shown in Fig. \ref{Fig.5} by the directed dashed lines. The presence of some small field here is necessary to remove the ferromagnetic instability of disordered states, without the field the inhomogeneous states will jump right to the ferro-state due to this instability.

The exhaustive study of both these processes was made in the graphite-intercalated magnet $Cu_{0.93}Co_{0.07}Cl_2 $ having the phase sequence of path B \cite{22}. The similarity of the experimental data to $\bar m$ shown in Figs. \ref{Fig.13}, \ref{Fig.5} is quite impressive as authors has revealed the existence of numerous metastable states using multiple cooling-heating cycles, see the example in Fig. \ref{Fig.13}. Also in Ref.[\onlinecite{23}] "thermal remagnetization" similar to that in Fig. \ref{Fig.13} was observed in disordered ferromagnets $Nd_3(Fe,Ti)_{29} $ and $Nd_3(Fe,Re)_{29} $. The depolarization and polarization on heating were also seen in PLZT ferroelectric ceramics \cite{24,25}.

At last we should mention the example of the relaxor the thermodynamics of which is definitely distinct from that of the present model. It is PMN-PT with lead titanate (PT) content between 0.06 and 0.2 in which the temperature evolution of hysteresis loop shows the gradual transition from the inclined to the rectangular form \cite{33} instead of the sharp one as in Fig. \ref{Fig.6}. This means that between glass and ferro-phase in this compound there are the sequence of mixed phases where the most stable states are those with partially ordered dipole moments. Probably the model with the three- and four - mode interactions can describe such intermediate mixed phases.

\section{Discussion and conclusions}
Thus the present thermodynamic theory can describe in a unified manner a wealth of nonergodic phenomena in disordered ferromagnets and ferroelectrics. It demonstrates what the effects of numerous metastable states can actually be and how they can manifest themselves in various regimes of real-time quasi-static experiments. Also it shows what a form the results of rigorous statistical mechanics should have to describe the phase transitions in crystals with random 
interactions.

The important conclusion can be made on the relations between the predictions of eqiulibrium thermodynamics valid in the infinite-time limit and finite-time experiments. Contrary to naive expectations that there are no such relations the present study shows that they can be revieled in finite-field quasi-static experiments. In particular, they are straightfordly manifested in the change of hysteresis loop form closely related to the form of the static thermodynamic potential. Yet its form and thermodynamic path for a given crystal can be obtained in some other sufficiently full set of finite-field quasi-static experiments. 'Sufficiently full' means here that a wide spectrum of inhomogeneous metastable states should be explored to determine the regions of their existence and their thermodynamic properties. The exellent examples of such studies providing the valuable data for the future theory are given in Refs. [\onlinecite{18}, \onlinecite{22}].

Apparantly the present phenomenology can be improved and expanded in many ways. Thus one may consider a multicomponent order parameter as, say, in Heisenberg ferromagnet or cubic ferroelectric and less sparse modes having three- and four-mode couplings in Landau's potential. Also one may include in it higher-order terms in $m_i$ to expand the results to larger $T$ and $h$ regions. Eventually it may be worth to explore the models with the broken in some way permutation symmetry. It seems also important to consider the role of fluctuations (i. e. the modes which do not condense) in the present approach. 
The quantitative description of the nonergodic thermodynamics of PMN - type relaxors needs also the consideration of random-fields.

\begin{acknowledgments}
I gratefully acknowledge useful discussions with M.P. Ivliev, I.P. Raevski, V.B. Shirokov, V.I. Torgashev, E.D. Gutlianskii, S.A. Prosandeev, V.P. Sakhnenko.
\end{acknowledgments}

\end{document}